# Bidirectional quantum teleportation and secure direct communication via entanglement swapping

Shima Hassanpour [a,*] and Monireh Houshmand [b]

[a] MS Student, Department of Electrical Engineering, Imam Reza International University, Mashhad, Iran
[b] Assistant Professor, Department of Electrical Engineering, Imam Reza International University, Mashhad, Iran

**Abstract**

In this paper, a bidirectional quantum teleportation protocol based on Einstein-Podolsky-Rosen (EPR) pairs and entanglement swapping is proposed. In this scheme, two users can simultaneously transmit an unknown single-qubit state to each other. The implementation of the proposed scheme is easier in experiment as compared to previous work. By utilizing this bidirectional quantum teleportation protocol, a bidirectional quantum secure direct communication scheme without carrying secret message is presented. Therefore, in the case of using perfect quantum channel, the protocol is completely secure. Finally, security analyses are investigated.

*Key words: Bidirectional quantum teleportation; bidirectional quantum secure direct communication; entanglement swapping; EPR pair.*

## 1. Introduction

Entanglement swapping is an important aspect of quantum information theory [1]. This feature is widely applied in quantum information processing tasks such as quantum teleportation (QT) [2], quantum secure direct communication (QSDC) [3], quantum secret sharing (QSS) [4] and so on.

Quantum teleportation is a process of transmitting an unknown quantum state from one place to another through shared entanglement and some auxiliary classical communications. A protocol for quantum teleportation (QT) was first introduced in 1993 by Bennett et al., [5]. Subsequently, several quantum teleportation protocols have been proposed by utilizing EPR pair [6], GHZ state [7], $W$ state [8 − 10], GHZ-like state [11 − 13] and cluster state [14, 15].

Recently, Fu et al., [16] presented a bidirectional quantum teleportation (BQT) scheme via four-qubit cluster state as a quantum channel. In their scheme, two users can simultaneously exchange their single-qubit states using Hadamard operation, defined unitary operations and Bell basis measurement.

Controlled quantum teleportation (CQT) is a type of QT in which a state is teleported under the permission of the controllers. In 1998 [17], Karlsson and Bourennane presented the first controlled quantum teleportation by using Greenberger-Horne-Zeilinger (GHZ) state. Accordingly, lots of protocols for CQT have been proposed [18 − 23].

Bidirectional quantum controlled teleportation (BQCT) as a new field of QT, has attracted a number of authors in recent years [24 − 29]. In a BQCT scheme Alice and Bob can simultaneously transmit an unknown quantum state to each other under the supervision of all the controllers.

The first protocol in BQCT was proposed by Zha et al., [24] in 2010. They utilize Brown state [30] for implementing their scheme. Three years later, Zha et al., [25] presented another scheme via five-qubit cluster state as a quantum channel. In 2013, Li and Nie [26] proposed a BQCT protocol by using a five-qubit composite GHZ-Bell state. In that year, Shukla et al., [31] proved that the permission of the controller in Li's and Nie's

---

* Corresponding author at: Department of Electrical Engineering, Imam Reza International University, Mashhad, Iran Tel.: +98-882-179-9.

E-mail address: shimahassanpour@yahoo.com.

scheme [26] is limited to one direction only. Also, they introduced a class of five-qubit quantum state that can be utilized for BQCT. In the same year Yan [27] presented a scheme via six-qubit cluster state.

In 2014, Duan et al [28] proposed a BQCT protocol by using maximally seven-qubit entangled state. Also, they mentioned because of utilizing seven-qubit state as quantum channel, the controller needs to perform measurement three times. Therefore, it can improve greatly the security of this scheme. In that year, Chen [29] presented another scheme via five-qubit entangled state. In all these BQCT schemes that have been introduced, Alice and Bob can send an unknown single-qubit state under the permission of the controller. Besides, no one has not illustrated their scheme is as a generalization of bidirectional quantum teleportation (BQT).

On the other hand, quantum secure direct communication (QSDC) is a distinct communication way to which secret message is directly sent through quantum channel without sharing a private key. In 2002 [32], Beige et al., proposed the first QSDC protocol by using single photons. Based on the idea of QSDC, bidirectional quantum secure direct communication (BQSDC) or the so-called quantum dialogue protocol was first proposed by Nguyen [33]. In the BQSDC protocol, two legitimate users can simultaneously exchange their secret messages. In addition to this scheme, a number of such protocols have been presented recently [34 − 37].

In this paper, a protocol of bidirectional quantum teleportation by using EPR pair and the property of entanglement swapping is presented in the first step. In the second step, a bidirectional quantum secure direct communication scheme via BQT is proposed. In the proposed BQSDC scheme, Alice and Bob as a two legitimate users can read out each one encoded message simultaneously. Besides, no qubits carrying the secret messages are transmitted in a quantum channel.

The rest of this paper is organized as follows. In the following section a protocol for BQT via EPR pair is provided. The comparison between proposed BQT and Fu et al.'s scheme [16] is given in Section III. In Section IV, it is shown that the proposed BQT protocol may be converted to a BQSDC protocol. In Section V, the security analyses of the proposed BQSDC protocol is investigated. Finally, the conclusions are provided in Section VI.

## 2. Proposed Bidirectional quantum teleportation protocol

Before detailing the proposed scheme, we briefly introduce entanglement swapping method. This method can entangle two particles that have not been directly interacted. To illustrate this method we consider EPR pairs that are maximally entangled states. The four EPR pairs are described as Eq. (1),

$$|\phi^+\rangle_{12} = \frac{1}{\sqrt{2}}(|00\rangle + |11\rangle)_{12}, \qquad |\psi^+\rangle_{12} = \frac{1}{\sqrt{2}}(|01\rangle + |10\rangle)_{12},$$
$$|\phi^-\rangle_{12} = \frac{1}{\sqrt{2}}(|00\rangle - |11\rangle)_{12}, \qquad |\psi^-\rangle_{12} = \frac{1}{\sqrt{2}}(|01\rangle - |10\rangle)_{12}. \qquad (1)$$

Suppose two users share $|\psi^+\rangle_{12}$ and $|\psi^+\rangle_{34}$ where the sender has qubits one and four and the receiver has qubits two and three. In this step, if the sender applies Bell state measurement on his qubits, the state of the whole system collapse and the receiver's qubits become entangled with each other as Eq. (2),

$$|\psi^+\rangle_{12} \otimes |\psi^+\rangle_{34} = \frac{1}{2}(|01\rangle + |10\rangle)_{12} \otimes (|01\rangle + |10\rangle)_{34}$$
$$= \frac{1}{2}(|01\rangle_{12}|01\rangle_{34} + |01\rangle_{12}|10\rangle_{34} + |10\rangle_{12}|01\rangle_{34} + |10\rangle_{12}|10\rangle_{34})$$
$$= \frac{1}{2}(|01\rangle_{14}|10\rangle_{23} + |00\rangle_{14}|11\rangle_{23} + |11\rangle_{14}|00\rangle_{23} + |10\rangle_{14}|01\rangle_{23})$$
$$= \frac{1}{2}(|\phi^+\rangle_{14}|\phi^+\rangle_{23} - |\phi^-\rangle_{14}|\phi^-\rangle_{23} + |\psi^+\rangle_{14}|\psi^+\rangle_{23} - |\psi^-\rangle_{14}|\psi^-\rangle_{23}). \qquad (2)$$

Therefore, if the sender's outcome is $|\psi^-\rangle_{14}$, the receiver's outcome will be $|\psi^-\rangle_{14}$. In fact, the state after each user's measurement becomes one of the EPR pairs that have maximally entangled state. Also, similar results can be achieved if other Bell state is shared by two users.

In the proposed protocol, we suppose Alice and Bob want to teleport a single-qubit state to each other. Their single-qubit state are described as Eq. (3),

$$|\emptyset\rangle_A = \alpha_0|0\rangle + \alpha_1|1\rangle, \qquad |\emptyset\rangle_B = \beta_0|0\rangle + \beta_1|1\rangle. \tag{3}$$

This scheme consists of the following steps:
Step1 Assume that the quantum channel linking Alice and Bob is a two-EPR entangled state, which has the form of Eq. (4),

$$|\emptyset\rangle_{a_1 b_1 a_2 b_2} = \frac{1}{2}(|00\rangle + |11\rangle)_{a_1 b_1} \otimes (|00\rangle + |11\rangle)_{a_2 b_2} = \frac{1}{2}(|0000\rangle + |0011\rangle + |1100\rangle + |1111\rangle)_{a_1 b_1 a_2 b_2}, \tag{4}$$

where the qubits $a_1, a_2$ belong to Alice and qubits $b_1, b_2$ belong to Bob. The state of the whole system can be expressed as Eq. (5),

$$|\Psi\rangle_{a_1 b_1 a_2 b_2 AB} = |\emptyset\rangle_{a_1 b_1 a_2 b_2} \otimes |\emptyset\rangle_A \otimes |\emptyset\rangle_B. \tag{5}$$

Step2 Alice and Bob make a CNOT operation with qubits $A$ and $B$ as control qubits and qubits $a_1$ and $b_2$ as targets respectively. The state will be the form of Eq. (6),

$$\begin{aligned}|\Psi'\rangle_{a_1 b_1 a_2 b_2 AB} = \frac{1}{2}[&(|0000\rangle + |0011\rangle + |1100\rangle + |1111\rangle)_{a_1 b_1 a_2 b_2} \alpha_0 \beta_0 |00\rangle_{AB} \\
&+ (|0001\rangle + |0010\rangle + |1101\rangle + |1110\rangle)_{a_1 b_1 a_2 b_2} \alpha_0 \beta_1 |01\rangle_{AB} \\
&+ (|1000\rangle + |1011\rangle + |0100\rangle + |0111\rangle)_{a_1 b_1 a_2 b_2} \alpha_1 \beta_0 |10\rangle_{AB} \\
&+ (|1001\rangle + |1010\rangle + |0101\rangle + |0110\rangle)_{a_1 b_1 a_2 b_2} \alpha_1 \beta_1 |11\rangle_{AB}]. \end{aligned} \tag{6}$$

Step3 Alice and Bob perform a single qubit measurement in the $Z$ basis on qubits $a_1$ and $b_2$ and the $X$ basis measurement on qubits $A$ and $B$ respectively. Then they announce their results to each other. As Table I shows, by applying appropriate operations each one can reconstruct the other's single-qubit state. As an example, if Alice's measurements result in the first step is $|0\rangle_{a_1}|+\rangle_A$, and Bob's measurements outcome is $|0\rangle_{b_2}|+\rangle_B$, the state of the remaining particles collapse into the state as Eq. (7),

$$|\Omega\rangle_{b_1 a_2} = (\alpha_0\beta_0|00\rangle + \alpha_0\beta_1|01\rangle + \alpha_1\beta_0|10\rangle + \alpha_1\beta_1|11\rangle)_{b_1 a_2} = (\alpha_0|0\rangle + \alpha_1|1\rangle)_{b_1}(\beta_0|0\rangle + \beta_1|1\rangle)_{a_2}. \tag{7}$$

Now, each legitimate user can reconstruct the single-qubit state by applying a suitable unitary operation as shown in Table I. Thus, the bidirectional teleportation is successfully finished.

## 3. Comparison

In this section, we compare the proposed BQT protocol with Fu et al.'s scheme [16]. Recently, Fu et al., [16] proposed a scheme of bidirectional teleportation via four-qubit cluster state as a quantum channel. In their scheme, two users can simultaneously find out the other one secret message by applying Hadamard operation, Bell bases measurement and defined unitary operations. But, in the proposed BQT protocol, as defined in Section II, we use two-EPR pairs as a quantum channel. Indeed, we utilize the property of entanglement swapping. Also, in contrast to Fu et al.'s scheme [16], we use single-qubit measurements in the proposed scheme.

To summarize, compared with Fu et al.'s scheme [16], the advantages of the proposed BQT protocol are described as follows:

1) The quantum channel between users is a two-EPR entangled state which is more easily prepared in the experiment compared to four-qubit cluster state [38]. Therefore, the proposed BQT protocol is easier to carry out.
2) The entanglement must be preserved between two qubits, while maintaining entanglement between four qubits is more complicated.
3) Single-qubit measurements are applied which are more efficient than Bell state measurements [39].

Table 1. Relation between measurement results and operations of users

| Measurement result | Operation |
|---|---|
| 0 + | $I$ |
| 1 + | $\delta^x$ |
| 0 − | $\delta^z$ |
| 1 − | $\delta^{iy}$ |

## 4. Modification of the BQT protocol to the BQSDC protocol

In this section, we illustrate the proposed BQT protocol can be generalized to a BQSDC protocol. The proposed BQSDC scheme is described in the following subsections:

*4.1. Preparation*

Alice prepares a large ordered number of identical two-particle EPR pairs as Eq. (8). For each pair she keeps the first particle with herself and sends the second particle to Bob.

$$|\emptyset^+\rangle_{ab} = \frac{1}{\sqrt{2}}(|00\rangle + |11\rangle)_{ab}. \qquad (8)$$

*4.2. Eavesdropping check*

After Bob acknowledges that he has received all of the second particles, Alice randomly selects sufficiently particles of her sequence as checking particles. To measure the checking particles, she randomly chooses one of the $Z$ or $X$ basis. Eq. (9) shows the representation of the EPR pair in the $X$ basis. Then, Alice announces the orders and the bases of the checking particles to Bob. Bob measures the corresponding particles under the same measuring bases. Then, he announces his measurement results to Alice. According to Table II, Alice controls the security of the quantum channel. If there is no Eve in the quantum channel, the measurement results must be completely correlated and they continue the communications, otherwise they discard it.

$$|00\rangle = \frac{1}{2}(|++\rangle + |+-\rangle + |-+\rangle + |--\rangle), \quad |11\rangle = \frac{1}{2}(|++\rangle - |+-\rangle - |-+\rangle + |--\rangle),$$
$$|\emptyset^+\rangle = \frac{1}{\sqrt{2}}(|00\rangle + |11\rangle) = \frac{1}{\sqrt{2}}(|++\rangle + |--\rangle). \qquad (9)$$

Table 2. Measurement bases and measurement results

| Basis | $\sigma_z$ | $\sigma_z$ | $\sigma_x$ | $\sigma_x$ |
|---|---|---|---|---|
| Alice's result | 0 | 1 | + | − |
| Bob's result | 0 | 1 | + | − |

*4.3. Encoding*

After ensuring the security of the quantum channel, Alice and Bob agree to divide all of the remaining EPR pairs into many groups. Each group contains two-EPR pair which can be used for Alice and Bob to exchange their one-bit secret message. Also, Alice and Bob produce a single-qubit state according to their one-bit secret message. The single state carrying secret message that Alice and Bob want to teleport to each other is represented as Eq. (10),

$$|\emptyset\rangle_A = \frac{1}{\sqrt{2}}(|0\rangle + \alpha|1\rangle)_A, \quad |\emptyset\rangle_B = \frac{1}{\sqrt{2}}(|0\rangle + \beta|1\rangle)_B, \qquad (10)$$

where $\alpha = 1$ ($\beta = 1$) and $\alpha = -1$ ($\beta = -1$) correspond to $|+\rangle$ and $|-\rangle$ respectively. Actually, $|+\rangle$ and $|-\rangle$ indicate the secret message, "0" and "1", respectively.

To achieve the purpose of bidirectional quantum secure direct communication, Alice and Bob send qubits $a_1, A$ and $b_2, B$ through a CNOT gate respectively as in step 2 of the previous protocol.

Now, Alice and Bob carry out a single-qubit measurement in the $Z$ basis on qubits $a_1$ and $b_2$ and perform a measurement in the $X$ basis on qubits $A$ and $B$ respectively. Then, they announce their measurement results to each other via the classical channel.

*4.4. Decoding*

Alice and Bob need to perform a defined operation on their qubit according to the classical bits that each one has received. Table I shows the suitable unitary operation. After performing unitary operations, Alice and Bob perform a measurement in the $X$ basis on qubit $a_2$ and $b_1$ respectively. Therefore, Alice and Bob can successfully read out the secret message. As an example, suppose Alice's and Bob's measurement results are $|0\rangle_{a_1}|+\rangle_A$ and $|0\rangle_{b_2}|+\rangle_B$ respectively. Then the state of the remaining qubits collapse in to the state as follows:

$$|\Omega\rangle_{b_1 a_2} = \frac{1}{8}[|00\rangle + \beta|01\rangle + \alpha|10\rangle + \alpha\beta|11\rangle]_{b_1 a_2} = \frac{1}{4}[\frac{1}{\sqrt{2}}(|0\rangle + \alpha|1\rangle)_{b_1} \otimes \frac{1}{\sqrt{2}}(|0\rangle + \beta|1\rangle)_{a_2}]. \quad (11)$$

According to Table I, Alice and Bob apply $I$ operation on their qubits. Now, Alice can read out Bob's and Bob can infer Alice's one-bit secret message by performing $X$ basis measurement on qubit $a_2$ and $b_1$ respectively. Therefore, the bidirectional quantum secure direct communication based on teleportation is successfully realized.

## 5. Security analyses of the proposed BQSDC protocol

In the proposed BQSDC scheme, transmission of secret message is inspired by teleportation. Therefore no qubits carrying secret messages are transmitted in the quantum channel. Actually, Eve may attack the quantum channel only in the preparation phase. In the following we will discuss some kinds of Eve's attack briefly. The first type of attack that is investigated is intercept-resend attack. In this attack, Eve measures Bob's qubit in the $X$ or $Z$ basis randomly and sends the new replacement qubit that she measured to Bob. Suppose Eve measures in the $Z$ basis. In this case, if the users basis in the checking process is $Z$, Eve's existence is not be detected as Eq. (8) shows and when the users basis is $X$, her interface is detected with $1/2$ probability as is described in Eq. (12). Now, suppose Eve measurers in the $X$ basis. In the case that the users measure in the $X$ basis, Eve's existence is not be detected as Eq. (9) shows and when the users basis is $Z$, according to Eq. (13), her interface will be detected with $1/2$ probability.

$$\frac{1}{\sqrt{2}}(|++\rangle + |--\rangle) = \frac{1}{2}(|+0\rangle + |+1\rangle + |-0\rangle - |-1\rangle). \quad (12)$$

$$\frac{1}{\sqrt{2}}(|00\rangle + |11\rangle) = \frac{1}{2}(|0+\rangle + |0-\rangle + |1+\rangle - |1-\rangle). \quad (13)$$

On the other hand, Eve can attack by applying a unitary operation as demonstrated in Eq. (14) on the Bob's qubit. Then she sends the qubit that she prepared (Eq. (15)) to Bob.

$$U = \begin{bmatrix} u_{00} & u_{01} \\ u_{10} & u_{11} \end{bmatrix}. \quad (14)$$

$$(I \otimes U)\frac{1}{\sqrt{2}}(|00\rangle + |11\rangle)_{ab} = \frac{1}{\sqrt{2}}(|0\rangle(u_{00}|0\rangle + u_{10}|1\rangle) + |1\rangle(u_{01}|0\rangle + u_{11}|1\rangle)$$
$$= \frac{1}{2\sqrt{2}}[|+\rangle(u_{00} + u_{01} + u_{10} + u_{11})|+\rangle + |+\rangle(u_{00} + u_{01} - u_{10} - u_{11})|-\rangle$$
$$+ |-\rangle(u_{00} - u_{01} + u_{10} - u_{11})|+\rangle + |-\rangle(u_{00} - u_{01} - u_{10} + u_{11})|-\rangle]. \quad (15)$$

According to Table I, the Eve's interface can be detected in the security checking process with probabilities of $\frac{1}{2}\{|u_{01}|^2 + |u_{10}|^2\}$ and $\frac{1}{8}\{|u_{00} + u_{01} + u_{10} + u_{11}|^2 + |u_{00} + u_{01} - u_{10} - u_{11}|^2 + |u_{00} - u_{01} + u_{10} - u_{11}|^2 + |u_{00} - u_{01} - u_{10} + u_{11}|^2\}$, if the users measurement bases are $Z$ and $X$ respectively.

## 6. Conclusions

In this paper, a theoretical bidirectional quantum teleportation protocol based on EPR pair and entanglement swapping is proposed. In the scheme, Alice may transmit an orbitary single-qubit state of qubit *A* to Bob and at the same time Bob also may send an orbitary single-qubit state of qubit *B* to Alice. In the scheme only single-qubit measurement, controlled-NOT operation and appropriate unitary operations are necessary.

Also, by utilizing this BQT, a bidirectional quantum secure direct communication scheme is presented in which none of the users could read out the secret message without co-operating the other one. In detail, after insuring the security of the quantum channel, Alice and Bob produce one auxiliary qubit according to their secret messages. After applying CNOT operation and single-qubit measurement, each one transmit two-bit classical information based on their measurement results to the other one. Only when the users collaborate, each one can find out the secret message correctly by applying appropriate unitary operation and single-qubit measurement. Because of utilizing entanglement swapping and teleportation technique, no qubits carrying secret messages are transmitted in a quantum channel. Therefore, this BQSDC protocol is unconditionally secure in the case of using perfect quantum channel.

In a future work, the proposed BQT protocol in this paper will be presented in order to teleport more than one qubit. In other words, two users can simultaneously transmit a particular type of two-qubit states to each other in a quantum channel.